\documentclass[preprint,epsf,showpacs]{revtex4}
\usepackage{amsmath}
\usepackage{epstopdf}
\usepackage{array}
\usepackage{hhline}
\usepackage{longtable}
\usepackage{epsfig}

\begin{document}

\title{Dynamic phase transitions on the kagome Ising ferromagnet}

\author{Zeynep Demir Vatansever}

\affiliation{Department of Physics, Dokuz Eyl\"{u}l
University, TR-35160, Izmir-Turkey}

\date{\today}
\begin{abstract}
We perform extensive Monte Carlo simulations to investigate the dynamic phase transition
properties of the two-dimensional kinetic Ising model on the kagome lattice in the presence
of square-wave oscillating magnetic field. Through detailed finite-size scaling analysis, 
we study universality aspects of the non-equilibrium phase transition. Obtained critical exponents indicate that the two-dimensional kagome-lattice kinetic Ising model belongs to the same universality class with the corresponding Ising model in equilibrium. Moreover, dynamic critical exponent of the local moves used in simulations is determined with high precision. Our numerical results are compatible with the previous ones on kinetic Ising models.
\end{abstract}

 \maketitle

\section{Introduction}
\label{sec:introduction}
When a typical ferromagnet is exposed to a time-dependent oscillating magnetic field below its Curie temperature, $T_C$, it can display dynamically ordered and disordered phases and a corresponding non-equilibrium dynamic phase transition (DPT)~\cite{Tome, Chakrabarti, Yuksel0}.
A basic model to study the dynamic phase transitions is the kinetic Ising model (KIM) which, despite its simplicity, enables us to reach the complex dynamics behind the non-equilibrium systems. KIM can represent a ferromagnetic system subjected to a time-dependent magnetic field, $h(t)$, with a half-period of, $t_{1/2}$, and amplitude, $h_0$. The competition between time scales of period of the external field and metastable lifetime of the system, which is defined as the average time for the system to pass through the state with zero magnetization,  leads to a dynamic phase transition at the critical period of the external field.  Such a symmetry breaking between dynamically disordered (paramagnetic) phase and ordered (ferromagnetic) phase was initially observed in a theoretical study of a mean-field model~\cite{Tome} and later in kinetic Monte Carlo simulations~\cite{Rao, Lo}.

Throughout the years, many theoretical and experimental studies have been devoted to understanding the physics behind DPTs~\cite{Chakrabarti, Yuksel0, Riego0, Zimmer, Acharyya1, Acharyya2, Acharyya3, Buendia1, Buendia2, Jang1, Shi, Punya, Riego}. Several analogies between thermodynamic and dynamic phase transitions, for instance, similar phase diagrams, have been shown in theoretical and experimental studies~\cite{Riego0, Berger, Ramirez, Quintana}. In addition, DPTs have been studied in a diversity of models such as nanoscale systems ~\cite{Yuksel1, Vatansever1, Wu}, systems with surfaces ~\cite{Riego, Park1, Tauscher, Aktas1, Aktas2}. 

In the last two decades, there has been a great effort towards to estimation of critical exponents and universality classes of spin systems exposed to a time-dependent
magnetic field. Successful implementations of finite-size scaling techniques on KIM have shown that thermodynamic 
and dynamic phase transitions belong to the same universality class for both 2D and 3D cases~\cite{Sides1,Sides2,Korniss,Buendia3, Vatansever2, Park2}. 
It is worth mentioning that these findings are also in agreement with the symmetry arguments of Grinstein et al.~\cite{Grinstein} and the study of Ginzburg-Landau model in an external oscillating field~\cite{Fujisaka}.
In Ref.~\cite{Vatansever_Fytas} the authors have found that the universality class of the Blume Capel (BC) model driven by a time-dependent magnetic field is the same as the equilibrium  BC model. KIM with a disorder in exchange interaction couplings (random-bond KIM)~\cite{Vatansever3} and crystal field 
strength (random-crystal-field KIM)~\cite{Vasilopoulos} have been the subject of recent studies and the authors have presented strong evidence that disordered 
KIM belongs to the same universality class as the equilibrium Ising model except for double-logarithmic corrections in the specific heat scaling behavior. 
The effect of surfaces on DPT has been analyzed by Park and Pleimling and remarkably, surface exponents in non-equilibrium case are reported to be different 
from equilibrium surface exponents~\cite{Park1}. Apart from the estimation of critical exponents, there are few attempts to provide information about the properties 
of critical dynamics of algorithms used in the simulations. For instance, the critical exponents of the 2D KIM have been estimated by using the standard Glauber and Metropolis dynamics and it has been shown that DPT is universal regarding to choice of the stochastic dynamics ~\cite{Buendia3}. 
Also, the autocorrelation function of the dynamic order parameter at the critical period shows the existence of critical slowing down which displays itself 
by increasing correlation times with increasing lattice sizes. Korniss et. al~\cite{Korniss} have determined the dynamic exponent for the Glauber single-spin-flip 
algorithm in KIM as $z=1.91(0.15)$ which is close to the dynamic exponent of 2D equilibrium Ising model with local dynamics ~\cite{Nightingale}.

Despite the above-mentioned attempts for the characterization of the universality class of KIM, 
there are still some unanswered points related to the critical properties of the model. For instance, 
critical exponents and the universality class of the KIM have not been determined on lattices which correspond to realistic materials.
Additionally, as far as we know, there has not been a precise estimate of the dynamic exponent $z$ of local dynamics used in the simulations for the systems subjected to a sinusoidal external drive except for the result for square-lattice KIM~\cite{Korniss}.
Therefore, the objective of the present work is to provide detailed estimates of the critical exponents of 2D KIM located on a kagome lattice, 
using extensive MC simulations based on the Metropolis algorithm and finite-size scaling tools. Kagome lattice is an appropriate model to represent the recent 
2D ferromagnetic materials that are promising 
candidates for the development of spintronic devices~\cite{Zheng}. In addition to the critical exponents, we have determined the dynamic 
exponent of single-spin-flip Metropolis algorithm at the dynamic phase transition. In a nutshell, it is possible to say that our estimate on the dynamic exponent is found to be very 
close to the dynamic exponent of 2D equilibrium counterpart supporting the previous estimate~\cite{Korniss}.

The remainder part of the paper is organized as follows: 
In Sec.~\ref{sec:model} we present the model and details of MC simulations.
Also, the thermodynamic quantities measured in the simulations are introduced.
Sec.~\ref{sec:results} includes our numerical results and discussion. 
Our conclusions are presented in Sec.~\ref{sec:conclusions}.

\section{Model and Simulation Details}
\label{sec:model}
In this study, we consider 2D kinetic Ising model located on a kagome lattice in the presence of a  
time-dependent oscillating magnetic field. The Hamiltonian of the system can be written as
\begin{equation}\label{eq:1}
 \mathcal{H}= -J\sum_{\langle xy \rangle}\sigma_{x}\sigma_{y}-h(t)\sum_{x}\sigma_{x},
\end{equation}
where the spin variable $\sigma_{x}$ takes the values $\{ \pm 1 \}$. $J>0$ represents 
the ferromagnetic exchange coupling constant and $\langle xy \rangle$ denotes the summation over
nearest-neighbor spins. The final term $h(t)$ represents periodically oscillating magnetic field
that is spatially uniform such that all lattice sites in the system are subjected to a square-wave magnetic field with amplitude $h_{0}$ and half period $t_{1/2}$~\cite{Korniss, Park1}. 
The time-dependent magnetization per site is given by
\begin{equation}\label{eq:2}
 M(t)=\frac{1}{N}\sum_{x = 1}^{N}\sigma_{x}(t),
\end{equation}
where $N$ is the number of total sites in the system. In order to observe dynamic phase transitions, we 
shall study various thermodynamic quantities with varying half-period of the external field. One of them is
dynamic order parameter which is the period-averaged magnetization~\cite{Chakrabarti, Yuksel0}
\begin{equation}\label{eq:3}
Q=\frac{1}{2t_{1/2}}\oint M(t)dt.
\end{equation}
Here, the integration is over one cycle of the oscillating magnetic field. 
Since the probability density of the order parameter is bimodal with two opposite peaks for such finite systems, we measure 
the average norm of the order parameter, $\langle |Q| \rangle$ in our calculations. 

In order to determine and characterize dynamic phase transition and also extract critical exponents using finite-size scaling tools, one has to
calculate the scaled variance of the dynamic order parameter which is analogous to the static susceptibility~\cite{Sides1, Sides2, Korniss}.
\begin{equation}\label{eq:4}
\chi_{L}^{Q} = N\left[\langle Q^2\rangle_{L} -\langle |Q|
\rangle^2_{L} \right],
\end{equation}
The usage of $\chi_{L}^{Q}$  has been confirmed as a proxy for the nonequilibrium susceptibility by fluctuation-dissipation relations ~\cite{Robb1}. 
In the same way, one can measure the scaled variance of the period-averaged energy
\begin{equation} \label{eq:5}
\chi_{L}^{E} = N\left[\langle E^2\rangle_{L} -\langle E
\rangle^2_{L} \right].
\end{equation}
$\chi_{L}^{E}$ can be considered as the relevant heat
capacity of the dynamic system. Here, $E$ denotes the
cycle-averaged energy corresponding to the cooperative part of the
Hamiltonian~(\ref{eq:1}). 

Moreover, we measure the fourth-order Binder cumulant
\begin{equation}\label{eq:6}
 U_{L}^{Q}=1-\frac{\langle Q^4\rangle_L}{3\langle Q^2
\rangle_L^2},
\end{equation}
via the dynamic order parameter $Q$ to determine the dynamic phase transition point~\cite{Binder81}.

\begin{figure}[t]
\centering
\includegraphics[width=10 cm]{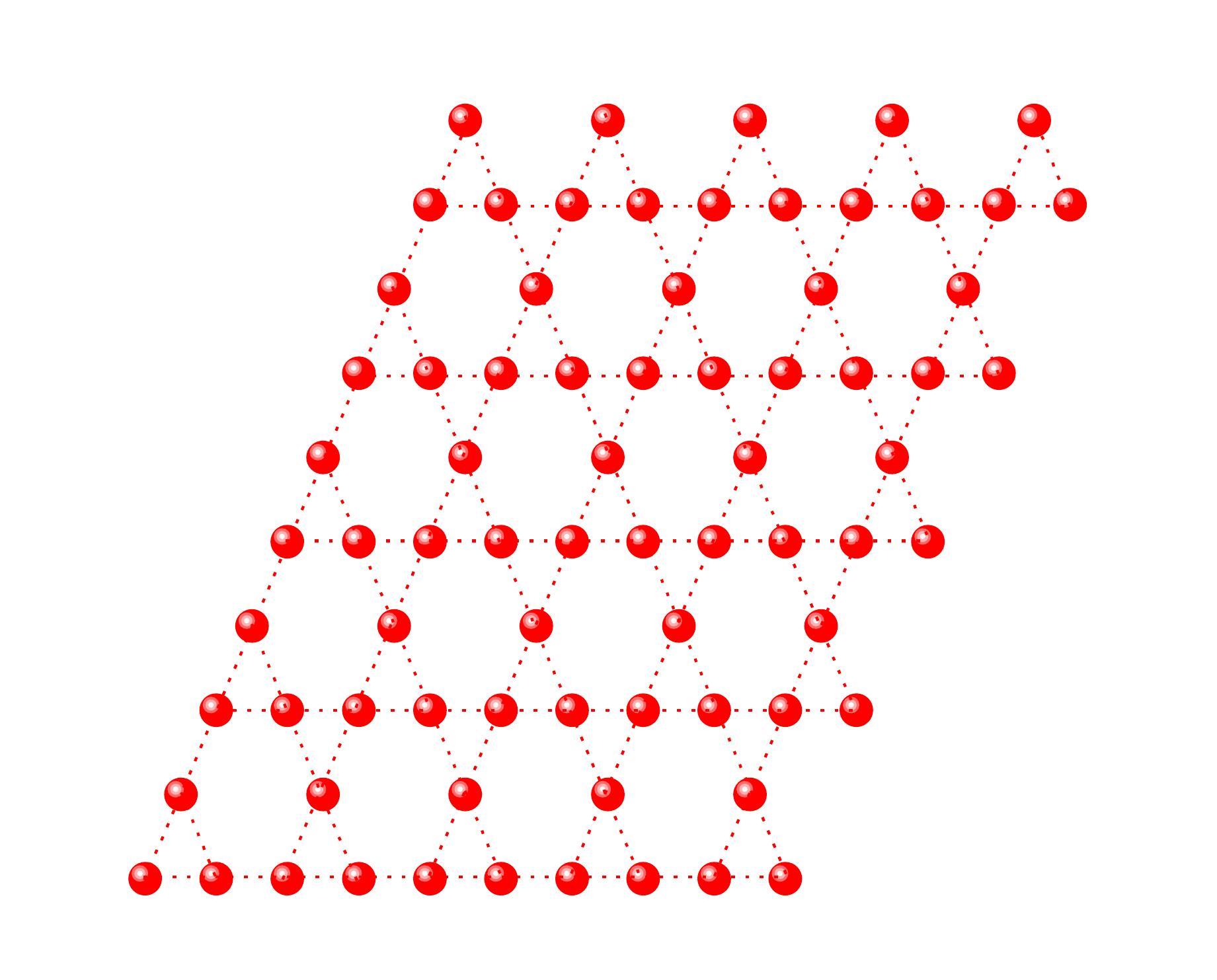}
\caption{\label{fig:kagomelattice} (Color online) Representation of kagome lattice with $L=5$. The lattice includes $3L^2$ sites.
}
\end{figure}

Monte Carlo simulations on the kagome lattice have been performed based on single-spin flip
Metropolis algorithm~\cite{Metropolis, Binder, Newman} which is proven to be successful 
in Kinetic MC simulations.  We carry out simulations on kagome lattices by updating the lattice sites randomly and enforcing helical boundary conditions. Kagome lattice is a regular array 
including hexagons and triangles (see Fig.~\ref{fig:kagomelattice}). Its unit cell contains three sites constructing an equilateral triangle~\cite{Newman}. Simulations are performed on lattices with $L \in \{15, 30, 45, 60, 75, 90, 105, 120, 150, 180, 210\}$ where $L$ is the dimension of the lattice in unit cells. The total number of lattice sites is $N=3L^2$. For the system to reach thermodynamic equilibrium, the first $2\hspace{0.05cm}500$ periods of the oscillating field have been discarded and thermal average of several quantities is calculated from next $25\hspace{0.05cm}000$ periods.
The unit of time in our simulations is Monte Carlo step per site (MCSS). For each lattice size, $100$ independent computer experiments are performed and error calculations have been carried out by using the jackknife method~\cite{Newman}.

It is known that metastable decay of the system in field-reversals depends on the temperature, field, and system size. DPT occurs in the multi-droplet (MD) regime where the metastable decay of the system takes place through the nucleation and growth of many droplets~\cite{Rikvold, Sides1}. In order to study in MD regime, the amplitude of the external square-wave magnetic field is chosen as $h_{0} = 0.3$ and temperature is fixed as $T = 0.8\times T_{\rm c}$ where $ T_{\rm c}$ is the critical temperature of the corresponding equilibrium model~\cite{Korniss}. The critical temperature of the 2D kagome lattice is available as $T_c=2.14332 J/k_B$ ($k_B$ is the Boltzmann constant) which has been calculated exactly by Sy\^{o}zi~\cite{Syozi}.

\begin{figure}[t]
\centering
\includegraphics[width=10 cm]{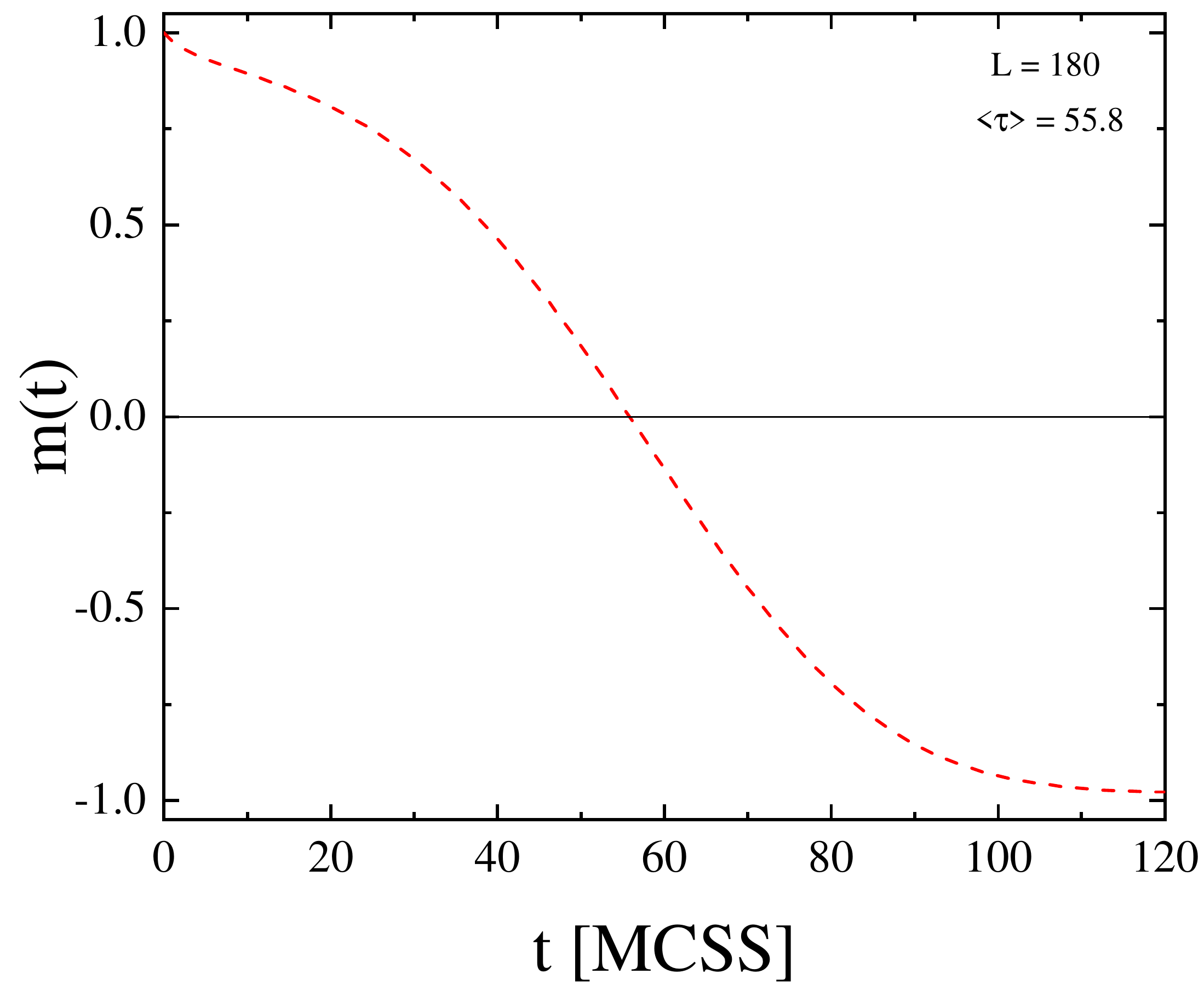}
\caption{\label{fig:m_vs_t} (Color online) Magnetization as a function of time at $T=0.8T_c$ for a lattice size of $L=180$. The system was initially 
in a state with all spins up. Then, a constant magnetic field in the opposite direction of the initial spin alignment is applied.}
\end{figure}

In order to measure the metastable lifetime, $\langle\tau \rangle$, we choose the initial configuration as a fully ordered state. When a constant magnetic field is applied in the opposite direction of the spin alignment at a temperature below the Curie temperature of the system, the magnetization changes by nucleating droplets that align in the same direction as the constant field. Determination of metastable lifetime is shown in Fig.~\ref{fig:m_vs_t} for a lattice size of $L=180$. Here, we have performed $25\hspace{0.05cm}000$ independent simulations and according to our simulations, metastable lifetime is determined as 
$\langle\tau \rangle=55.8$ (in terms of MCSS) at $T=0.8T_c$ for the considered system.
Metastable lifetime has been also determined for various lattice sizes 
and it has been found that $\langle\tau \rangle$ is independent of the system size in agreement with earlier results~\cite{Park2}.

Throughout the paper, we present the thermodynamic quantities as a function of the competition parameter, which is defined as the ratio of the half-period of the external field to metastable lifetime 
$\langle\tau \rangle$
\begin{equation}\label{eq:7}
 \Theta=\displaystyle \frac{t_{1/2}}{\langle\tau \rangle},
\end{equation}
which is analogous to temperature in thermodynamic phase transitions.

In addition to the above thermodynamic quantities, we measure other useful quantities in order to provide 
further insight about the properties of dynamics used in the simulations. One of them is the time-displayed autocorrelation
function of the order parameter at $n$th period which is defined as ~\cite{Buendia3, Korniss}
\begin{equation}\label{eq:8}
C_L^Q (n)=\displaystyle \frac{\langle Q(i) Q(i+n)\rangle - \langle Q(i) \rangle^2}{\langle Q^2(i) \rangle-\langle Q(i) \rangle^2},
\end{equation}
where $Q(i)$ is the value of the order parameter at $i$th period. By benefiting from the autocorrelation function, we also calculate integrated 
correlation time ~\cite{Newman},
\begin{equation}\label{eq:9}
 \tau^{Q}_L=\displaystyle \frac{1}{2}+\sum_{n=1}^{\infty} C_L^Q (n),
\end{equation}
which enables us to calculate the dynamic exponent $z$. 

We use the data for $L \geq L_{\rm min}$ for the employment of finite-size scaling laws. In our fittings, we apply the standard $\chi^2$ goodness of fit test. In order to obtain an acceptable fit, we consider a fit as being acceptable if the probability $Q$ values are $10\% \le Q \le 90\% $.
\section{Results and discussion}
\label{sec:results}
In this section, critical properties of 2D KIM located on the kagome lattice have been presented. 
It is now well-established that finite-size scaling tools can be implemented for the non-equilibrium models~\cite{Korniss, Buendia3, Park2, Vatansever2, Vatansever_Fytas, Vatansever3, Vasilopoulos}. Accordingly, we carried out large-scale MC  simulations and use finite-size scaling tools to extract the critical exponents of the present system. 

\begin{figure}[t]
\centering
\includegraphics[width=10 cm]{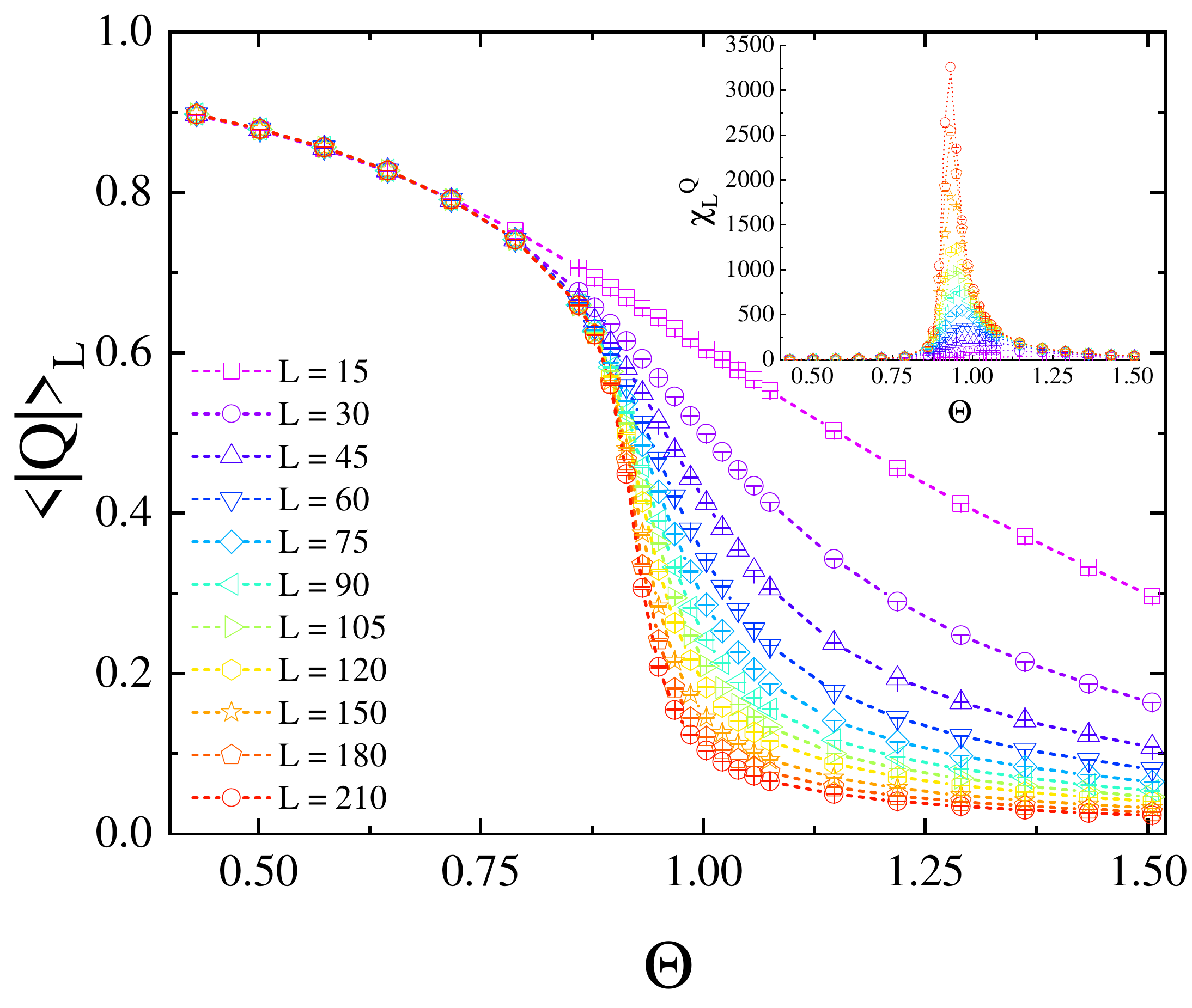}
\caption{\label{fig:Q_vs_Theta} (Color online) (Main panel) Dynamic order parameter as a function of $\Theta$ for various system sizes $L$. 
The inset shows the finite-size scaling behavior of the corresponding dynamic susceptibility $\chi_{L}^{Q}$. The error bars are smaller than the symbol sizes.}
\end{figure}

\begin{figure}[t]
\centering
\includegraphics[width=10 cm]{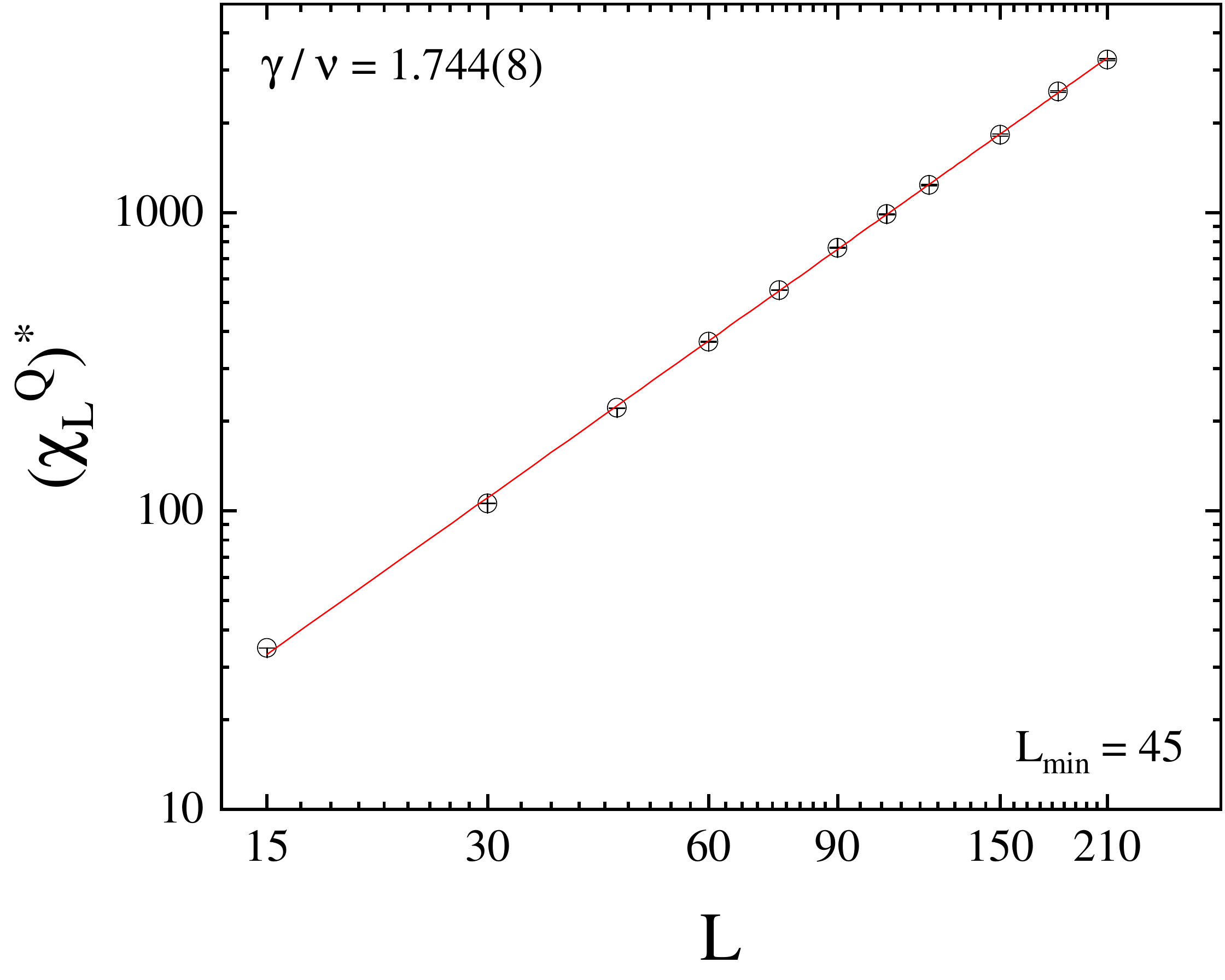}
\caption{\label{fig:Xq_vs_L} (Color online) Dynamic susceptibility maxima $(\chi_{L}^{Q})^*$ as a function of lattice size. The results 
are shown in log-log scale. The error bars are smaller than the symbol sizes.}
\end{figure}

We start our analysis by showing the dependency of the dynamic order parameter (main panel)
and the corresponding dynamic susceptibility (inset) on the competition parameter in Fig.~\ref{fig:Q_vs_Theta}. The order parameter takes finite values corresponding to a robust dynamically ordered phase for small half-period values whereas it approaches zero as  $\Theta$ increases. Apart from the small system sizes, the dynamic susceptibility has a divergent behavior and a typical peak near the phase transition indicating the existence 
of second-order phase transition in the system. This characteristic peak takes larger values with increasing system size. Fig.~\ref{fig:Xq_vs_L} displays the maxima of dynamic susceptibility, $(\chi^{Q}_{L})^{\ast}$, as a function of system size. The solid line is fit of the form
\begin{equation} \label{eq:10}
 (\chi^{Q}_{L})^{\ast} \sim L^{\gamma/\nu}.
\end{equation}
Based on our numerical data, the exponent is found as $\gamma/\nu=1.744(8)$ with a good agreement with the 2D Ising universality class value of $\gamma/\nu=7/4$.

\begin{figure}[t]
\centering
\includegraphics[width=10 cm]{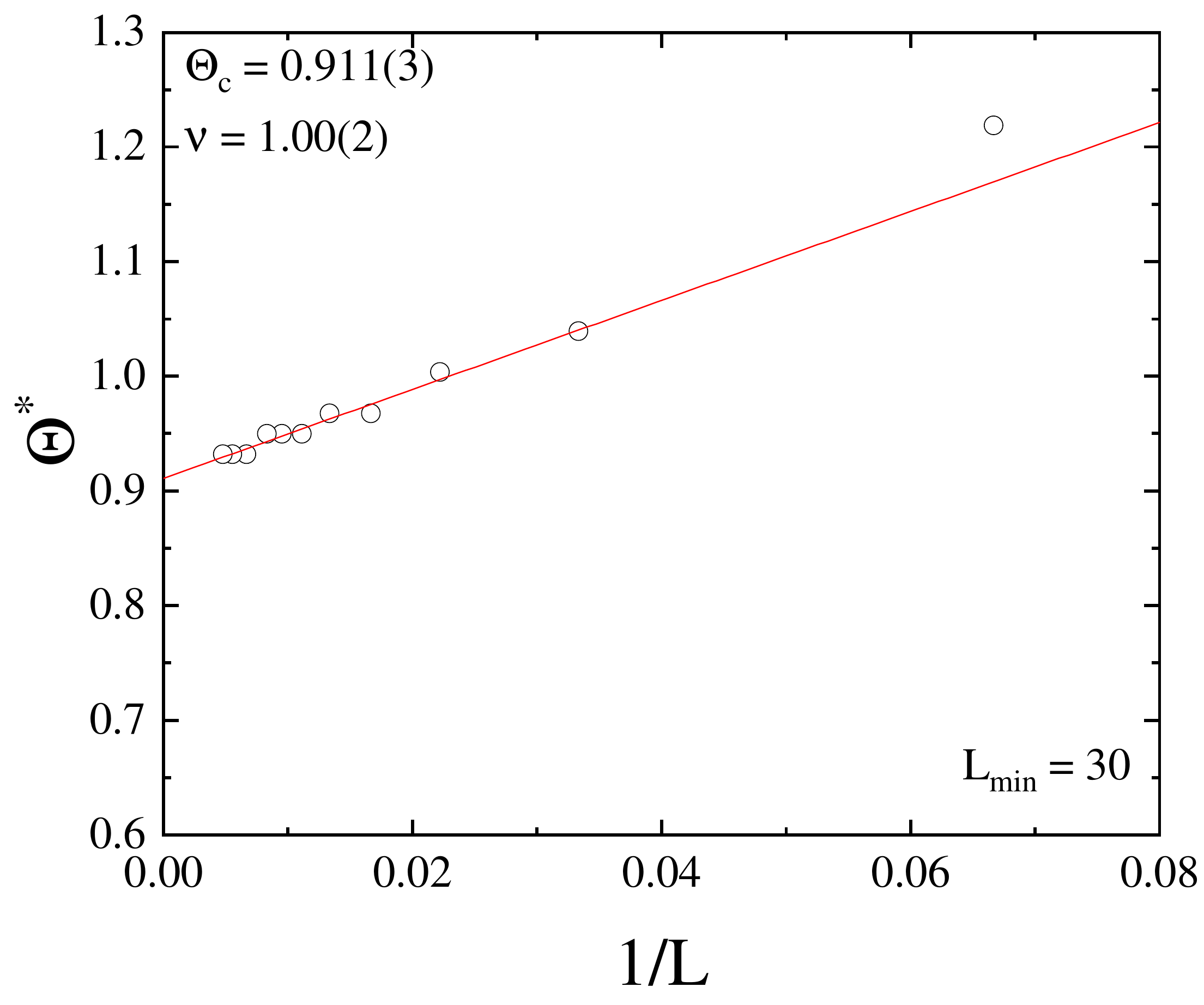}
\caption{\label{fig:Critical_Theta} (Color online) Shift behavior of pseudocritical competition parameter $\Theta^*$ obtained from the maxima of
dynamic susceptiblity.}
\end{figure}

\begin{figure}[t]
\centering
\includegraphics[width=10 cm]{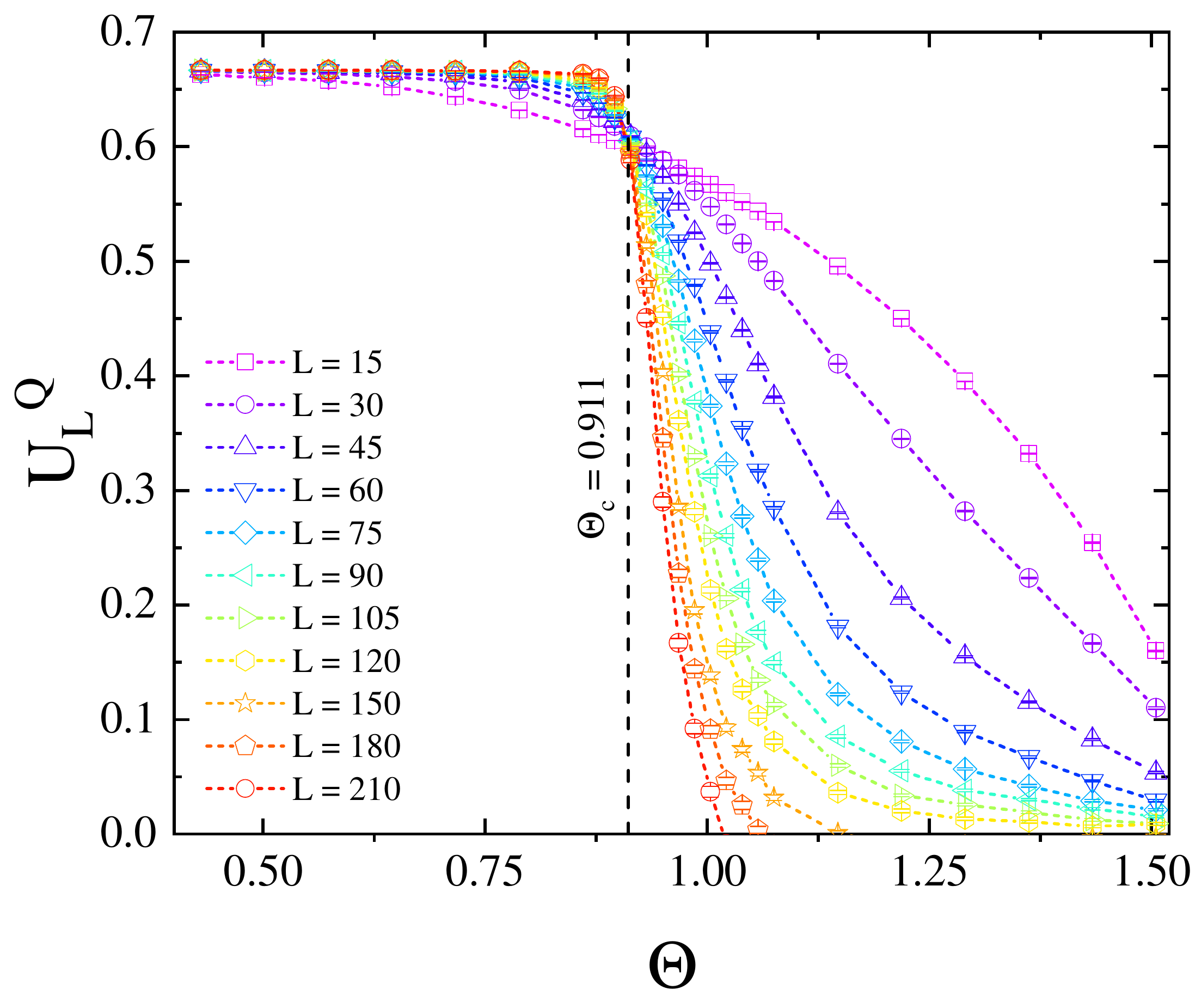}
\caption{\label{fig:Binder_vs_Theta} (Color online) The competition parameter dependency of fourth-order Binder cumulant $U_L^Q$ for various lattice sizes. The vertical dashed line implies a critical value of
$\Theta_c=0.911$. The error bars are smaller than the symbol sizes.}
\end{figure}

The locations of the peaks of the dynamic susceptibility, $\Theta^{\ast}$, obtained in finite-size systems can be used to estimate the critical competition parameter $\Theta_c$, at which an infinite system undergoes a phase transition, with a relation given below ~\cite{Vatansever_Fytas}
\begin{equation} \label{eq:11}
\Theta^{\ast} = \Theta_{\rm c} + bL^{-1/\nu}.
\end{equation}
Fig.~\ref{fig:Critical_Theta} shows $\Theta^{\ast}$ as a function of system size. The solid line is a fit of the form of Eq.~\ref{eq:11} which provides the critical competition parameter of $\Theta_c=0.911(3)$. This value is very close to unity implying that the DPT takes place
when the metastable lifetime of the system is comparable with the half-period of the external field. Also, the critical exponent of the correlation length is estimated as $\nu=1.00(2)$ with a clear agreement with the value $\nu=1$ of 2D Ising universality class ~\cite{Fisher, Privman, Binder, Fytas}.
In addition to the shift-behavior technique, we use the intersection method of the fourth-order Binder cumulant of the order parameter, $U_L^Q$, to determine $\Theta_c$ accurately~\cite{Binder81}. We present $U_L^Q$ defined in Eq.~\ref{eq:6} as a function of $\Theta$ for various lattice sizes in Fig.~\ref{fig:Binder_vs_Theta}. The vertical dashed line in the figure indicates a critical value of $\Theta_c=0.911$ in agreement with our analysis of Fig~\ref{fig:Critical_Theta}. Also, the obtained $\Theta_c$ values are compatible with the peak position of the response function 
$\chi^{Q}_{L}$ illustrated in Fig~\ref{fig:Q_vs_Theta}.

Estimation of the exponent of order parameter $\beta/\nu$ is performed by determining the order parameter at the positions of $(\chi^{Q}_{L})^{\ast}$ for all the system sizes considered as shown in Fig.~\ref{fig:Q_vs_L}.
The scaling behavior
\begin{equation} \label{eq:12}
 \langle |Q| \rangle _L \sim  L^{-\beta/\nu}
\end{equation}
helps us to estimate critical exponent as $\beta/\nu=0.125(5)$. This value is again very close to $\beta/\nu=1/8$ of  2D equilibrium Ising model within errors ~\cite{Fisher, Privman, Binder, Fytas}. 

\begin{figure}[t]
\centering
\includegraphics[width=10 cm]{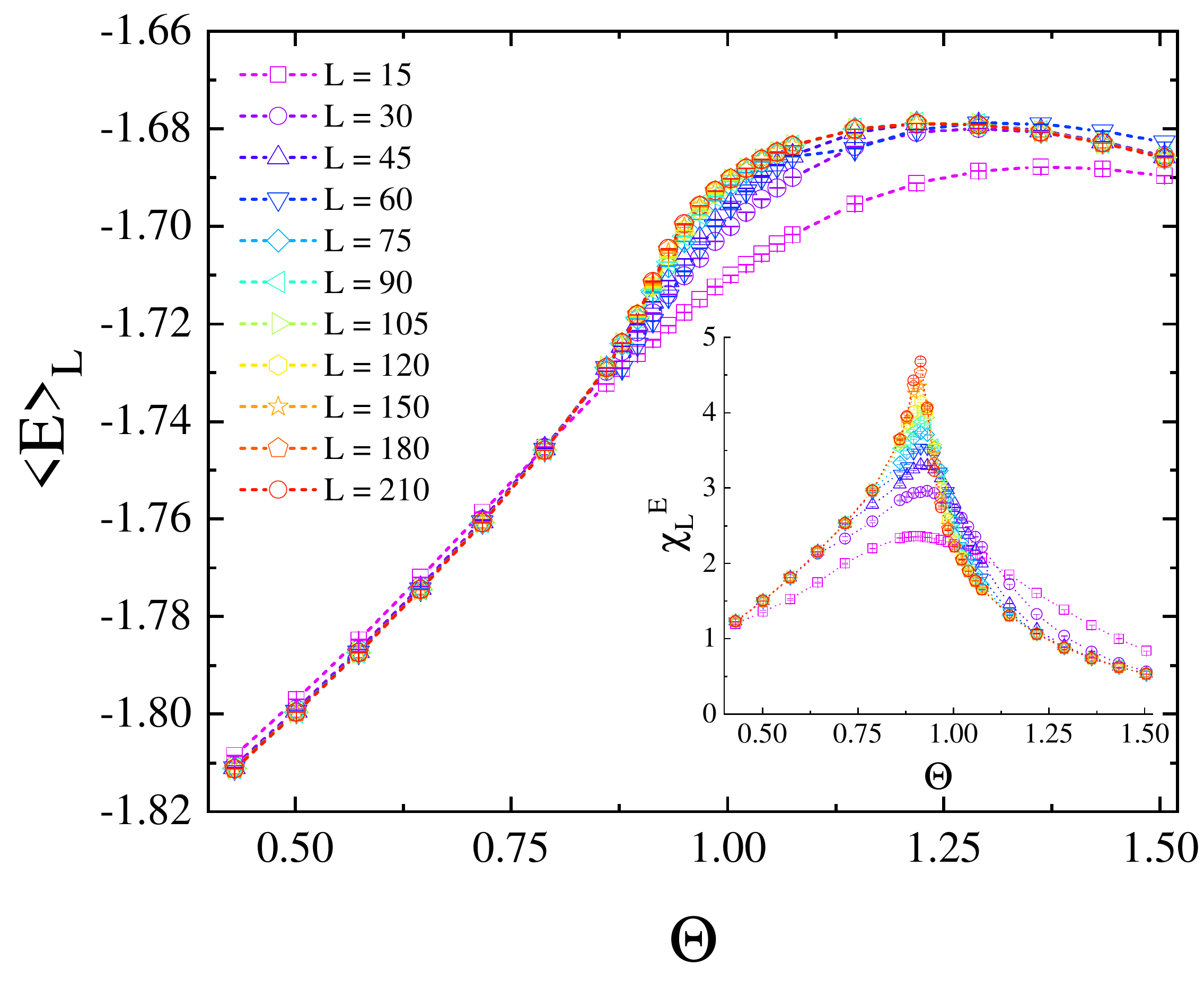}
\caption{\label{fig:E_vs_Theta} (Color online) The competition parameter dependency of period-average internal energy for various lattice sizes. The 
inset shows finite-size behavior of the corresponding scaled energy variance $(X_L^E)$. The error bars are smaller than the symbol sizes.}
\end{figure}

\begin{figure}[t]
\centering
\includegraphics[width=10 cm]{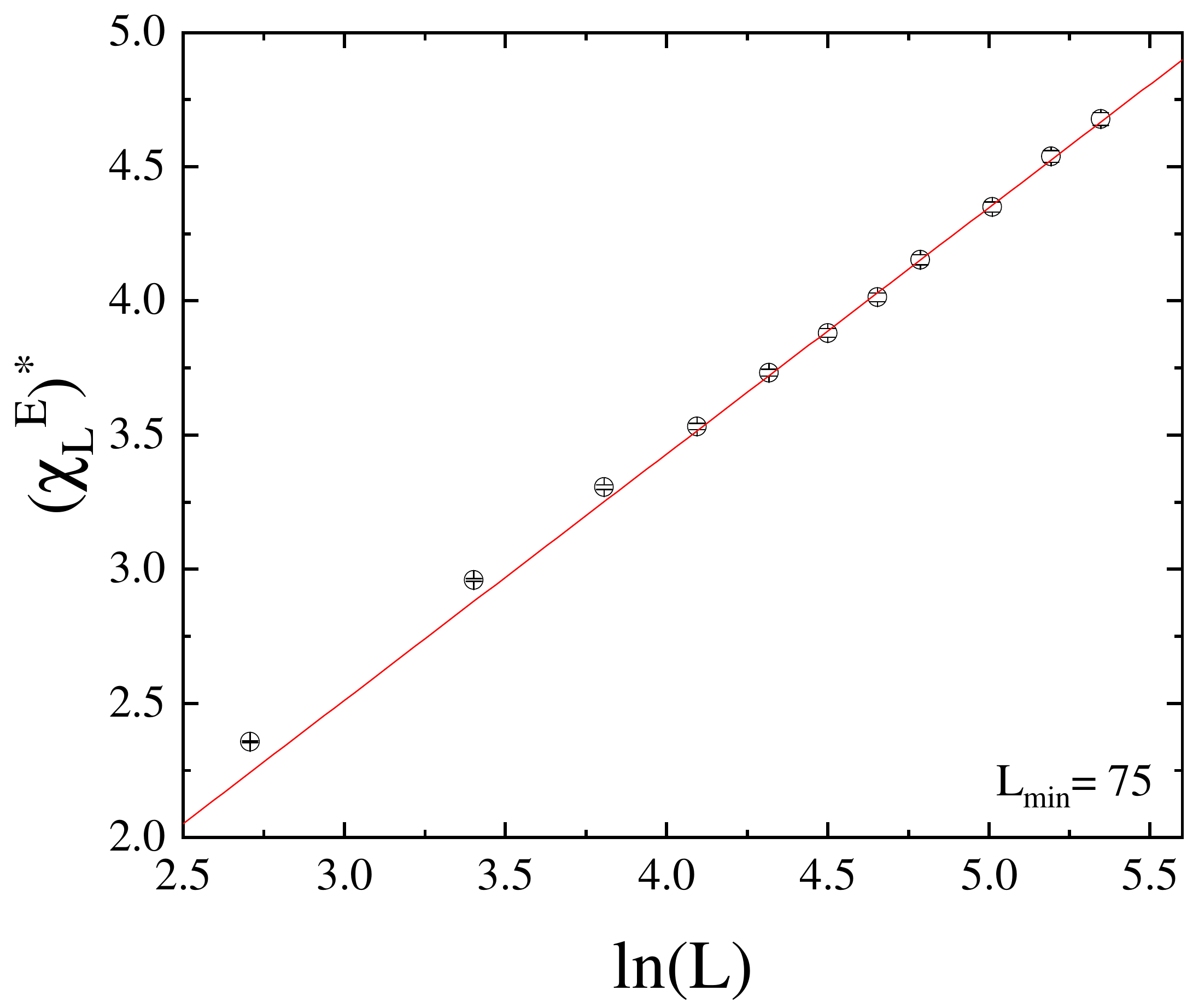}
\caption{\label{fig:Xe_vs_L} (Color online) Finite-size-scaling behavior of the maxima of scaled energy variance $(X_L^E)^\ast$. The error bars are smaller than the symbol sizes.}
\end{figure}

In order to give a complete description of universal aspects of the present system, we continue our finite-size scaling analysis by considering period averaged internal energy and the corresponding scaled variance which are displayed in the main panel and inset of Fig.~\ref{fig:E_vs_Theta}, respectively. A slow increment in the scaled variance of energy with increasing system size can be explicitly observed from the figure. 
Moreover, it is expected to observe a logarithmic scaling behavior of the
maxima of the heat capacity $(\chi_L^E)^*$ if the specific-heat critical
exponent $\alpha = 0$. Variation of maxima of scaled energy variance
$(\chi_L^E)^*$ as a function of system size in log-lin scale is
demonstrated in Fig.~\ref{fig:Xe_vs_L}. The numerical data represented here to show a clear logarithmic
divergence of the form~\cite{Ferdinand}
\begin{equation} \label{eq:111}
(\chi_L^E)^* \propto c_1 +c_2 \ln(L)
\end{equation}
as it is for the equilibrium Ising universality class. 

It is possible to say that the overall critical exponents estimated above are in good agreement with the previous results for KIM on two-dimensional lattices~\cite{Korniss, Buendia3, Vatansever2, Vatansever_Fytas, Vatansever3, Vasilopoulos}. Accordingly, one can conclude that the universality properties in DPT are independent of the topology of lattice. Our results together with the earlier ones in the literature imply that non-equilibrium phase transitions in KIM fall into the same universality class with its equilibrium counterpart, except for the systems including surface~\cite{Park1}. 

\begin{figure}[t]
\centering
\includegraphics[width=10 cm]{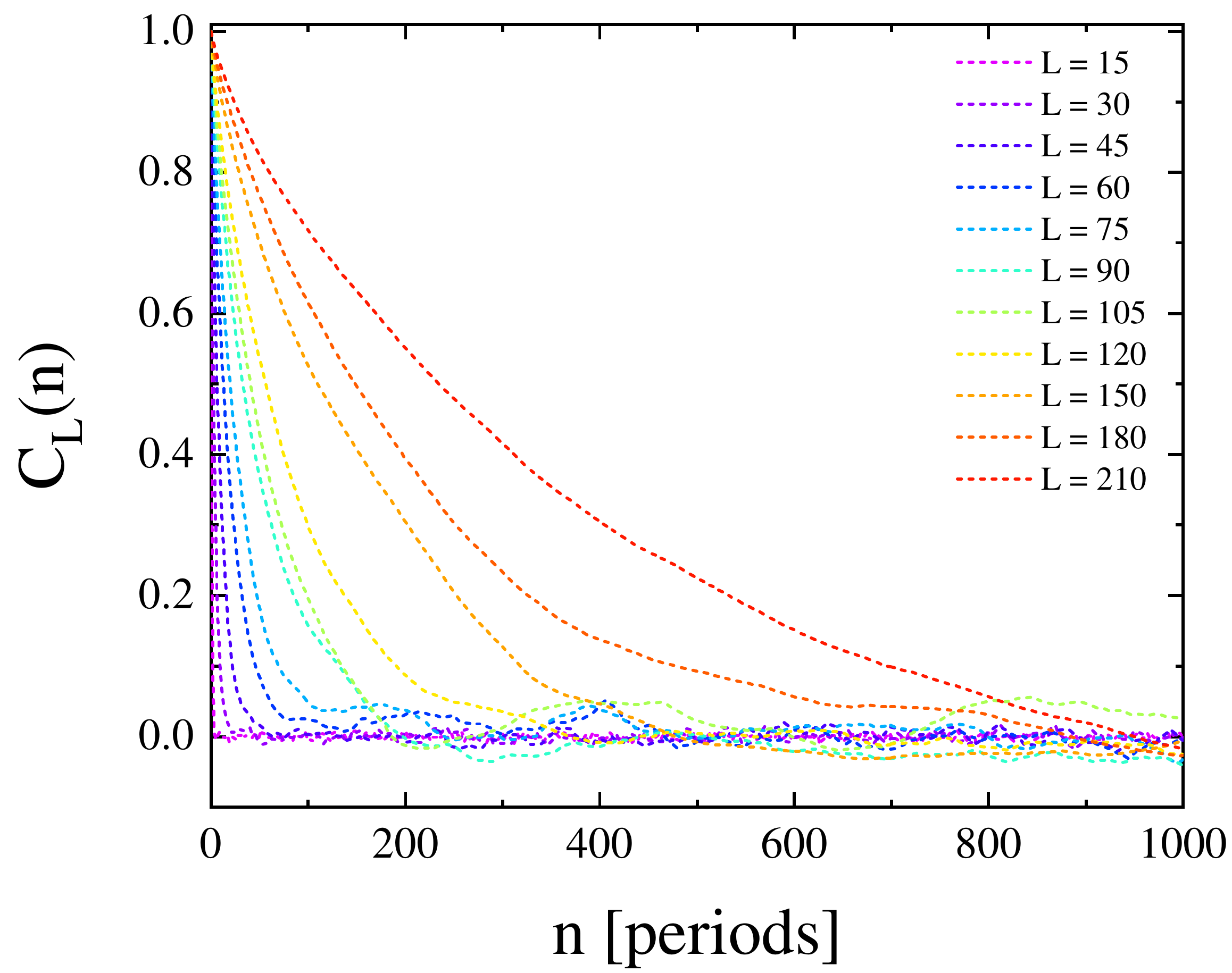}
\caption{\label{fig:C_vs_Period2} (Color online) Period dependence of normalized time-displaced autocorrelation function of the order parameter,
$C_L(n)$, at the critical point $\Theta_c$.}
\end{figure}

\begin{figure}[t]
\centering
\includegraphics[width=10 cm]{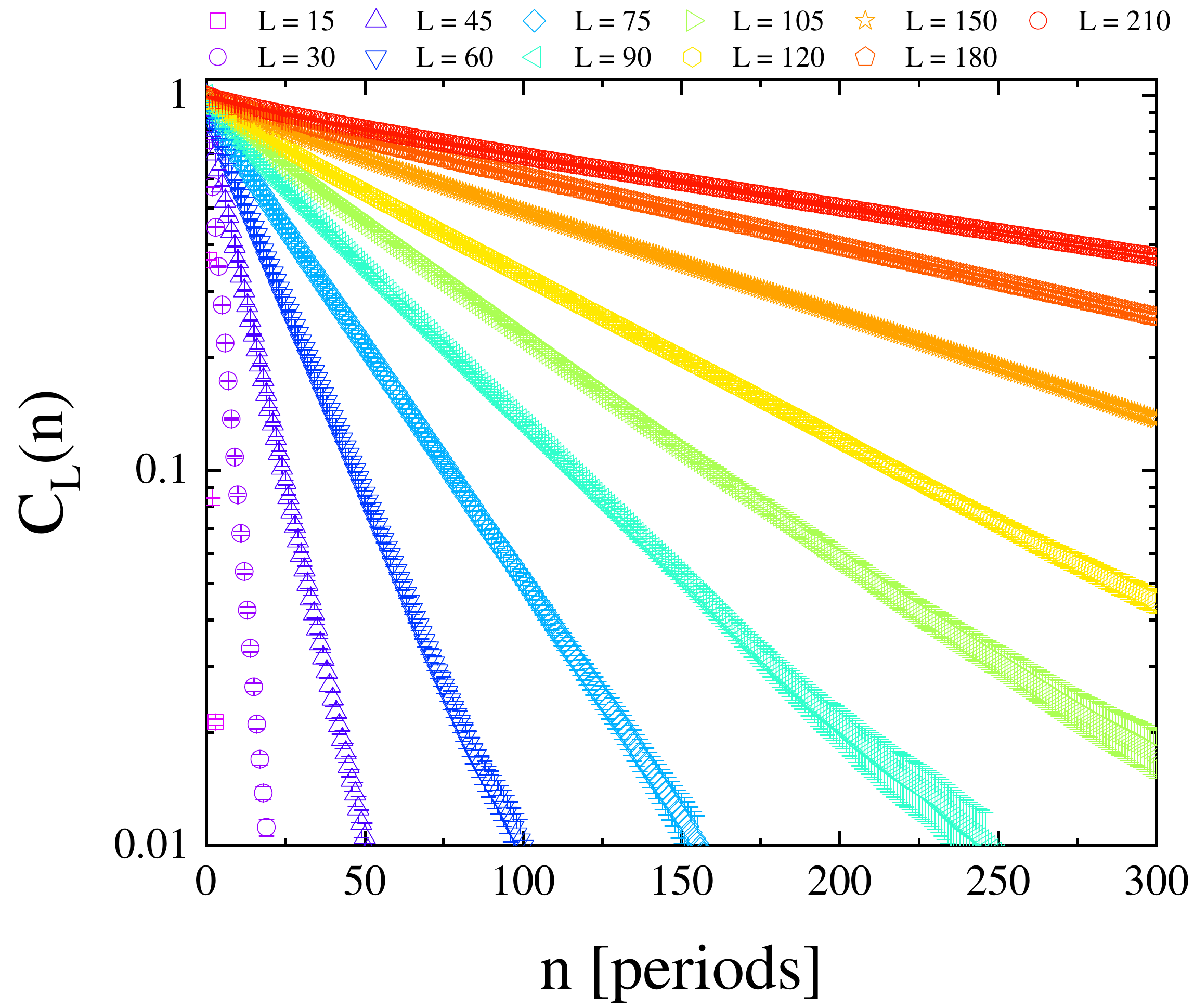}
\caption{\label{fig:C_vs_Period} (Color online) Time dependence of normalized time-displaced autocorrelation function for the order parameter on lin-log scale 
at the critical point $\Theta_c$. }
\end{figure}

\begin{figure}[t]
\centering
\includegraphics[width=10 cm]{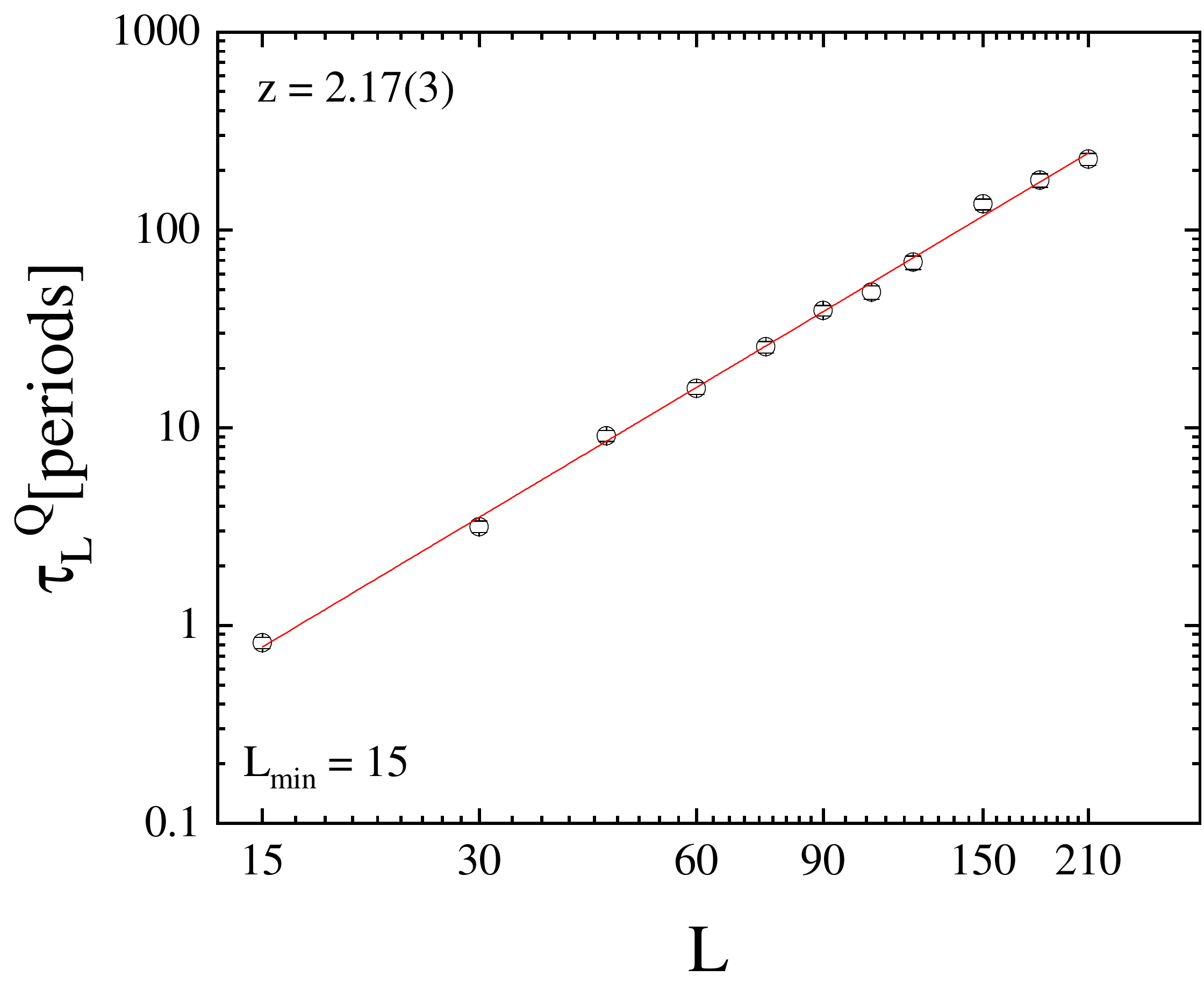}
\caption{\label{fig:Tau_vs_L} (Color online) Estimation of critical exponent $z$.
Integrated correlation time as a function system size is shown in log-log scale. The error bars are smaller than the symbol sizes.}
\end{figure}

Having the determined critical value of $\Theta_c$, it is worthwhile 
to study the details about the characteristics of dynamics used in our MC simulations.
The period dependency of time-displaced normalized autocorrelation function for magnetization defined in Eq.~\ref{eq:8} at the critical point, $\Theta_c$, is shown in Fig.~\ref{fig:C_vs_Period2}. The increment in correlation time with system size can be explicitly observed. This also indicates the existence of critical slowing down as anticipated in systems evolve under local Metropolis moves. Correlation time is expected to decay exponentially as:
\begin{equation}\label{eq:13}
C_L (n) \sim \exp({-n/\tau_L^{Q,Exp}}), 
\end{equation}
which can be checked by plotting the autocorrelation function in lin-log scale as shown in Fig.~\ref{fig:C_vs_Period}. Here, $\tau_L^{Q,Exp}$ is the exponential correlation time.
The correlation time gets higher with increasing $L$. The curves show that there is a clear critical slowing down in the system at the critical competition parameter. An alternative way to provide insight about characteristics of simulations is to calculate integrated correlation time, $\tau_L^Q$, defined in Eq.~\ref{eq:9}. Lattice size dependence of $\tau_L^Q$ is depicted in Fig.~\ref{fig:Tau_vs_L}. 
The correlation time is expected to obey ~\cite{Newman}
\begin{equation}\label{eq:14}
 \tau_L^Q \sim L^z
\end{equation}
at the critical point. A fit of the form of Eq.~\ref{eq:14} gives a dynamic exponent as $z=2.17(3)$. Precise value of the dynamic exponent of the 2D equilibrium Ising model obtained with single-spin flip dynamics is available as $z= 2.1665(12)$~\cite{Nightingale}. To the best of our knowledge, there is no detailed estimation of the dynamic critical exponent for KIM simulated with single-spin flip dynamics
except for the result reported by Korniss et al. when the system evolves under Glauber dynamics as $z=1.91(0.15)$~\cite{Korniss} on the square lattice. Therefore, combining the previous result in literature~\cite{Korniss} and our $z$ value, one may conclude that the dynamic critical exponent for single-spin flip dynamics in KIM is compatible with its equilibrium counterpart despite the presence of a time-dependent oscillating magnetic field.

\section{Conclusions}
\label{sec:conclusions}
In the present work, we studied the universality properties of KIM in two dimensions by extensive MC simulations. We particularly considered kagome lattice subjected to a periodic square-wave magnetic field below the Curie temperature of the system. By benefiting from finite-size scaling tools, we determined the critical exponent of the correlation length $\nu$, critical exponents ratios of magnetic susceptibility $\gamma/\nu$ and magnetization $\beta/\nu$ with high accuracy. The critical competition parameter at which a DPT occurs was obtained. Also,
a logarithmic divergence in finite-size behavior of the specific heat was observed. Obtained numerical results were found to be compatible with the critical exponents reported for 2D square-lattice KIM ~\cite{Korniss}. We additionally studied properties of the local dynamics used in MC calculations at the critical competition parameter. The dynamic exponent value was found as $z=2.17(3)$ which is comparable with that of equilibrium 
Ising model in 2D ~\cite{Nightingale} confirming the previous result obtained for square-lattice KIM~\cite{Korniss}.

In summary, our numerical findings indicate that 2D KIM on the kagome lattice belongs to the 
same universality class as 2D equilibrium Ising model. Despite the fact that our knowledge 
about critical phenomena in equilibrium systems have been well established, the same is not the case for non-equilibrium systems since there are very limited studies regarding the dynamic critical exponent of the model systems under the presence of a time-dependent magnetic field. It is worth noting that further studies are needed to have a better understanding of DPT and its relevant critical properties. Therefore, we believe that this work contributes for the classification of universal properties of non-equilibrium systems and trigger further theoretical and experimental studies in the field.

\begin{acknowledgments}
The numerical calculations reported in this paper were performed
at T\"{U}B\.{I}TAK ULAKBIM (Turkish agency), High Performance and
Grid Computing Center (TRUBA Resources).
\end{acknowledgments}

\newpage

\begin{figure}[t]
\centering
\includegraphics[width=10 cm]{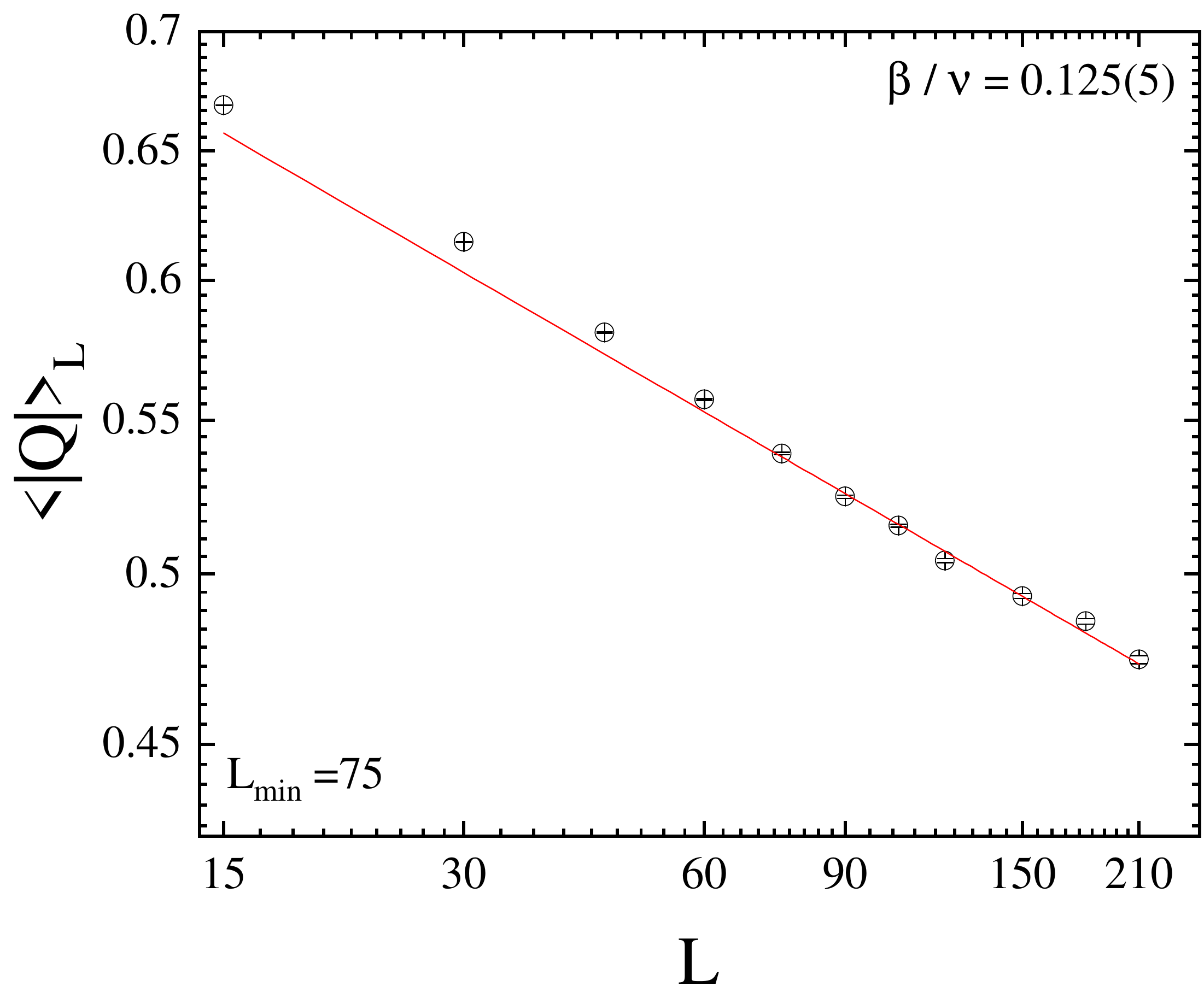}
\caption{\label{fig:Q_vs_L} (Color online) Order parameter at the positions $(\chi_{L}^{Q})^{\ast}$ as a function of system size in log-log scale. The error bars are smaller than the symbol sizes.
\label{fig:Q_vs_L} }
\end{figure}

\end{document}